# A dynamic resource allocation decision model for IT security

**Lotfi Hajjem[1], Salah Benabdallah[2], Fouad Ben Abdelaziz[3]**

[1] Graduate Student, Institut Supérieur de Gestion (ISG), University of Tunis, Tunis, Tunisia
(lotfi.hajjem@gmail.com)
[2] Director of Institut Supérieur des Etudes Technologiques en Communications de Tunis (Iset'Com), Tunis, Tunisia (sba@supcom.rnu.tn)
[3] Professor, College of Engineering, American University of Sharjah, Sharjah, United Arab Emirates.
(fabdelaziz@aus.edu)

**ABSTRACT**

Today, with the continued growth in using information and communication technologies (ICT) for business purposes, business organizations become increasingly dependent on their information systems. Thus, they need to protect them from the different attacks exploiting their vulnerabilities. To do so, the organization has to use security technologies, which may be proactive or reactive ones. Each security technology has a relative cost and addresses specific vulnerabilities. Therefore, the organization has to put in place the appropriate security technologies set that minimizes the information system's vulnerabilities with a minimal cost. This bi-objective problem will be considered as a resources allocation problem (RAP) where security technologies represent the resources to be allocated. However, the set of vulnerabilities may change, periodically, with the continual appearance of new ones. Therefore, the security technologies set should be flexible to face these changes, in real time, and the problem becomes a dynamic one. In this paper, we propose a harmony search based algorithm to solve the bi-objective dynamic resource allocation decision model. This approach was compared to a genetic algorithm and provided good results.

## 1. INTRODUCTION

Securing the information system is an important task for an organization. In fact, the continued growth in using information and communication technologies (ICT) for business purposes makes business organizations increasingly dependent on their information systems. Any successful attack will cause a serious loss of data, services, assets, business operations, etc. [7]. These attacks, which can be made by internal or external entities, exploit the vulnerabilities that may exist in the information system. To face these attacks, the organization has to overcome the information system's vulnerabilities using security technologies. Each one of the security technologies addresses specific vulnerabilities. Therefore, the organization has to put in place the appropriate set of security technologies that minimizes the vulnerabilities of the information system. This problem can be stated as a resource allocation problem (RAP). A RAP is the process of allocating resources among various projects or business units with a maximum profit and a minimum cost [1]. In the proposed model, the security technologies represent the resources to be allocated to the information system to overcome its vulnerabilities.

However, the set of vulnerabilities may change periodically with the continual appearance of new ones. Therefore, the set of security technologies needs to be flexible to face these changes. Thus, the problem, here, becomes a dynamic one, as the set of implanted security technologies should be redefined in real time to face the new vulnerabilities appearing in the information system. As a result, the studied problem will be stated as a dynamic resource allocation problem (DRA).

In addition, in this problem we have to consider the cost of each security technology. Thus, the organization wants to minimize the overall cost of the security technologies used to secure its information system. The problem becomes a bi-objective one, where we have, to minimize the number of vulnerabilities in the information system with the minimum cost.

This paper will be stated as follows: the problem of information system security will be described, in the first section. Next, in section 3, the problem of IT security, stated as a bi-objective dynamic resource allocation problem, will be defined and formulated. Then, the harmony search approach will be presented. Section 5 will be devoted to the adaptation of two resolution approaches, the harmony search algorithm and the genetic algorithm. And in the last section, a comparison of the two approaches is described. The paper finishes by a conclusion.

## 2. INFORMATION SYSTEM SECURITY

### 2.1 Why securing information systems?

The continued growth in the use of information technologies for business purposes makes business organizations increasingly dependent on their information systems. In fact, the evolution of network technologies permits an easy communication between different partners, independently of their locations. The communication may expose the partner's information assets to dangerous threats that exploit the information system vulnerabilities. An information asset is defined as anything of value to the organization. It can be either tangible or intangible. Tangible assets include physical infrastructure (such as servers and network infrastructure) and software elements of the information system. Intangible assets include business or other digital information of value to the organization (such as banking transactions, interest calculations, product-development plans and specifications), organization knowledge, company reputation and the intellectual property stored within the organizational system [7].

As it can be seen, the assets are of great importance for the organization. However, they are exposed to multiple threats that can be either natural disasters or human acts. The threats caused by human can be non-malicious (e.g., missing security patches, opening a malicious email, etc.) or malicious ones (e.g., theft, loss or destruction of an organizational asset, unauthorized access to the network services, infection with malicious code, insider threats, hackers, terrorists, etc.). Therefore, the organizations need to secure their information system against the threats that may exploit a large number of vulnerabilities. In fact, it is reported that the security research community identifies and publishes on an average of 40 new security vulnerabilities per week on various products, from operating systems, databases, applications to even networking devices [2]. Another study of the Computer Emergency Response Team/Coordination Center (CERT/CC) indicates that the number of found vulnerabilities was from 345 to 5990 in the decade of 1996 − 2005 [5]. Due to the large number of vulnerabilities, the number of attacks is growing in an immeasurable way. In fact, the number of events reported to CERT/CC was 2573 in 1996. In 2003, it was in an astonishing number of 137529 security incidents.

**2.2 Information security technologies**

Securing the information systems becomes a priority for the organizations. In fact, any successful attack on the information system and its eventual crash could result in a serious loss of data, services and business operations. Therefore, the organizations need to protect their information systems against the eventual attacks that may occur. To do so, they need to use efficient information security technologies that permit the protection of information and minimize the risk of exposing it to unauthorized parties. There are two families of security technologies, proactive and reactive ones. A proactive information security technology is a technique that takes preventative measures in a bid to secure data or resources before a security breach can occur [8] (e.g. cryptography, digital signature, virtual private network, etc.). Whereas, a reactive information security technology performs preventive measures in a bid to secure data or resources as soon as a security breach is detected [8] (e.g. firewalls, passwords, intrusion detection systems, etc.).

Each one of the security technologies addresses specific vulnerabilities and has a relative cost. Thus, the organization has to put in place the appropriate set of security technologies that minimizes the information system's vulnerabilities with the minimum cost, which becomes a big dilemma for it. In fact, according to the Department of Trade and Industry (DTI) 2006 survey the average is around 4 to 5% of the organization's IT budget being spent on security solutions [7].

In this paper, the problem of securing information systems will be stated as a Bi-objective Dynamic Resource Allocation Problem. This decision model will be defined, in the next section, and its mathematical formulation will be described.

**3. DECISION MODEL**

**3.1 Problem definition**

The problem of securing information systems will be studied, in this paper, as a RAP where the security technologies represent the set of resources. The problem can be stated as follows: Let $V$ be the set of vulnerabilities of an information system where:

$$V_i = \begin{cases} 1, \text{if vulnerability i is used by the information system} \\ \text{of the organization,} \\ 0, \text{Otherwise.} \end{cases}$$

Let $S$ be a set of $M$ security technologies that may be put in place by an organization where:

$$S_j = \begin{cases} 1, \text{if security j is used by the information system of the} \\ \text{organization,} \\ 0, \text{Otherwise.} \end{cases}$$

Each security technology $s_j$ has an associated cost $C_j$. And let $SV$ be the security/vulnerability matrix such that:

$$SV[i,j] = \begin{cases} 1, \text{if vulnerability i is covered by security j,} \\ 0, \text{Otherwise.} \end{cases}$$

The problem, here, is to find the set of security technologies that minimizes the number of vulnerabilities of the information system with the minimal cost. This problem will be studied as a bi-objective one where we

have to:
(1) Minimize the number of vulnerabilities at any time period $t$,
(2) Minimize the total cost of the security technologies to be used.

In addition, the problem will be studied dynamically in order to be able to overcome the new vulnerabilities that may appear, at each time. In fact, the organization needs to adapt the set of implanted security technologies, in real time, to the different circumstances that may happen to the information system, and with the minimal cost.

## 3.2 Mathematical formulation

The problem of securing information systems consists on finding the 'optimal' combination of security technologies that minimizes the information system's vulnerabilities, with the minimal cost. As defined in the last paragraph, it is a bi-objective problem that will be stated as dynamic resource allocation problem. Let $R$ be the set of residual vulnerabilities, where $r_i$ is calculated as follows:

if $v_i = 0$ then $r_i = 0$
if $v_i = 1$ and $\exists\ j/\ s_j = 1$ and $SV[i, j] = 1$, then $r_i = 0$.
if $v_i = 1$ and $SV[i, j] = 0\ \forall\ j|\ s_j = 1$, then $r_i = 1$.

Where the first condition indicates that if vulnerability $v_i$ is not present in the organization's information system ($v_i=0$) then it is not a residual vulnerability. The second one signifies that if vulnerability $v_i$ is present in the organization's information system ($v_i=1$) and there exists a security technology $s_j$ used by the organization that addresses it then it is not a residual vulnerability ($r_i = 0$). And the last equation signifies that if vulnerability $v_i$ is present in the organization's information system ($v_i=1$) and there is no used security technology $s_j$ addressing it then it is a residual vulnerability ($r_i=1$).

Therefore, the problem can be formulated as follows:

$$Min\ N_t(R_t) = \sum_{i=1}^{n_v} r_{it}, t = 1,...,T \qquad (1)$$

$$MinC(S) = \sum_t C_t(S_t) = \sum_{j=1}^{n_s}\sum_t c_j * s_{jt} \qquad (2)$$

Subject to

$$\sum_t q_{mt} \le q_m, m = 1,...,M, \qquad (3)$$

$$f(R_{t+1}, IS_t, S_{t+1}) = IS_{t+1} \qquad (4)$$

$$r_{it} \in \{0,1\}, i = 1,...,n_v, t = 1,...,T \qquad (5)$$

$$s_{jt} \in \{0,1\}, j = 1,...,n_s, t = 1,...,T \qquad (6)$$

Where $n_v$ and $n_s$ are the numbers of vulnerabilities and security technologies, respectively. $T$ represents the number of time periods. $R_t$ is the set of residual vulnerabilities at time period $t$. $IS_t$ is the set of implanted security technologies. And, $S_t$ is the set of security technologies that may be used by the organization. In this formulation, equations (1) represent the objective functions of minimizing the set of residual vulnerabilities at each time period $t$. Next, equation (2) is the objective function of minimizing the overall cost of the security technologies. Then, equations (3) are the resources satisfaction constraints. Finally, equation (4) indicates that the set of implanted security technologies at the $(t+1)^{th}$ time period is a function of the set of residual vulnerabilities at the $(t+1)^{th}$ time period ($R_{t+1}$), the set of implanted security technologies at the $(t)^{th}$ time period ($IS_t$), and the set of security technologies that may be used by the organization at the $(t+1)^{th}$ time period ($S_{T+1}$). For this problem, a new time period has to be considered where at least a new vulnerability is detected in the organization's information system.

This problem was not well-studied in the literature and few approaches were developed for some other problems close to it. Among them, we can note a genetic algorithm that was proposed to solve the static bi-objective resource allocation problem [4]. In addition, some metrics for quantifying an ICT security investment are described in [7].

## 4. HARMONY SEARCH ALGORITHM

### 4.1 Algorithm description

The harmony search (HS) algorithm is developed to imitate the musician behavior trying to improve its musical harmony practice after practice using the set of the pitches played by each instrument. This process can be compared to the one of optimizing an objective function iteration by iteration using the values assigned for decision variables [6].

The HS algorithm includes five steps: parameters initialization, the harmony memory (HM) initialization, the new harmony improvisation, the harmony memory update and the check of termination criterion [3].

### 4.2 Step 1: Parameters initialization

In this step, the optimization problem is specified:
Minimize (or Maximize) $f(x)$; $x_i \in X_i$, $i=1, 2, ..., N$ where:
- $f(x)$ is an objective function
- $x$ is the solution vector composed of decision variables $x_i$
- $X_i$ is the set of possible values for decision variable $x_i$
- $X_i = \{x_i(1), x_i(2), ..., x_i(K)\}$ for discrete variables
- $N$ is the number of decision variables
- $K$ is the number of possible values for each discrete variable

The algorithm parameters are also specified during this step such as:
- The harmony memory size($HMS$): is the number of solution in the memory

- The harmony memory considering rate (*HMCR*); 0 ≤ HMCR ≤ 1; his typical values range from 0.7 to 0.99
- The pitch adjustment rate (*PAR*); 0 ≤ PAR ≤ 1; his selected values range is from 0.1 to 0.5
- Improvisations number or objective functions number

### 4.3 Step 2: Harmony memory initialization

During this step, *HMS* solutions are randomly generated to form the harmony memory. Each decision variable (xi) selects a value from its corresponding list (Xi). Then the fitness values are calculated for the generated solutions (equation 7).

$$\begin{bmatrix} x_1^1 & x_2^1 & \cdots & x_{N-1}^1 & x_N^1 & | f(X^1) \\ x_1^2 & x_2^2 & \cdots & x_{N-1}^2 & x_N^2 & | f(X^2) \\ . & \cdots & \cdots & \cdots & & | \cdots \\ . & \cdots & \cdots & \cdots & & | \cdots \\ . & \cdots & \cdots & \cdots & & | \cdots \\ x_1^{HMS-1} & x_2^{HMS-1} & \cdots & x_{N-1}^{HMS-1} & x_N^{HMS-1} & | f(X^{HMS-1}) \\ x_1^{HMS} & x_2^{HMS} & \cdots & x_{N-1}^{HMS} & x_N^{HMS} & | f(X^{HMS}) \end{bmatrix} \quad (7)$$

### 4.4 Step 3: New harmony improvisation

In this step, a new harmony vector is generated from the HM based on memory considerations, pitch adjustments, and randomization, as shown in equation 8:

$$x'_i = \begin{cases} x'_i \in \{X_i^1, X_i^2, ..., X_i^{HMS}\}, \text{with probability HMCR} \\ x'_i \in X_i, \text{with probability (1 - HMCR)} \end{cases} \quad (8)$$

Where, HMCR (harmony memory consideration rate) is the probability of choosing a value from the solutions stored in the HM. While (1- HMCR) is the probability of randomly choosing one feasible value from the set of all possible values for the corresponding decision variable.
While improvising the new harmony, each value chosen from HM is examined to determine whether it should be pitch-adjusted. This procedure uses the PAR parameter that sets the rate of adjustment for the pitch chosen from the HM as follows:

$$\text{Pitch adjusting decision for } x'_i = \begin{cases} Yes, \text{with probability PAR,} \\ No, \text{with probability (1 - PAR).} \end{cases} \quad (9)$$

The value of *(1 - PAR)* sets the rate of doing nothing. If the pitch adjustment decision for $x'_i$ is YES, $x'_i$ is replaced as follow:

$$x'_i \leftarrow x'_i \mp rand(\ ) * bw \quad (10)$$

where *bw* is an arbitrary distance bandwidth and *rand()* is a random number between 0 and 1 or between -1 and 1.

### 4.5 Step 4: Harmony memory update

If the new harmony vector is better than the worst harmony in the HM, judged in terms of the objective function value, the new harmony is included in the HM and the existing worst harmony is excluded from the HM.

### 4.6 Step 5: Termination criterion check

If the stopping criterion is satisfied, computation is terminated. Otherwise, Steps 3 and 4 are repeated. The stopping criteria may be either maximum number of improvisations or a maximum number of iteration without improvement of the solution.

## 5. RESOLUTION APPROACHES

To solve the problem of securing, in real time, an information system against the different attacks that may happen, two meta-heuristics were developed, a harmony search algorithm and genetic algorithm. These two approaches are composed of two phases, the static and the dynamic one. The static phase is applied for the initial system state *(t=0)*. And the dynamic one is applied to face the new vulnerabilities that may be found in the system. It should take into consideration the current security plan and the new security technologies that may appear.

### 5.1 Harmony search algorithm

The proposed HS algorithm for the static phase can be described as follows:
- Step1. Parameters initialization: The improvisations number is equal to 2, as the studied problem is a bi-objective one. In addition, the harmony memory will contain the non-dominated solutions and its size (HMS) will be set to 50. The rates HMCR and PAR will be set to 95% and 30%, respectively.
- Step2. Harmony memory initialization: In this step, 50 different solutions will be randomly generated. The solutions generation process will be as follows: the security technologies will be randomly selected one by one until a construction-stopping criterion is verified, i.e. the total cost exceeds a value *Cmax* or the number of residual vulnerabilities becomes less than a bound Nvmin. In order to get a better solution quality, a security technology is added only if it covers at least a residual vulnerability. In addition, and while generating the HM a new security/vulnerability matrix, noted *SV'* will be constructed. It will present for each security technology the vulnerabilities that it covered, effectively in the system. That is, for any security technology chosen in the construction process, the vulnerabilities that were covered by it will be recorded.
- Step3. New harmony improvisation: In this step, a new harmony is generated based on the HMCR and PAR rates. The generation process can be described as follows: A residual vulnerability is randomly selected and according to the HMCR value, a security technology will be chosen either from $SV'$ or from $SV$, i.e:

$$x' \text{ is selected from} \begin{cases} SV', \text{ with probability HMCR} \\ SV, \text{ with probability (1 - HMCR)} \end{cases} \quad (11)$$

Then, each time a security technology is selected from $SV'$, a pitch adjustment is performed with a probability $PAR$. It consists on selecting a security technology from the ones that was not applied for the current vulnerabilities and if it does not exist, the selection will be done among the ones addressing the current vulnerability.
- Step4. Harmony memory update: The generated solution will be added to the HM if it is not dominated by any existing solution. In addition, if it is added to the HM, all the solutions dominated by the new solution will be eliminated.
- Step5. Termination criterion check: The Steps 3 and 4 are repeated until there is no improvement of the HM for 50 successive iterations. And the solutions of the HM will constitute the set of non-dominated solutions. It contains the solutions that can be adapted to the current information system.

The proposed HS for the dynamic phase differs from the one of the static phase in Step2, the harmony memory initialization. In fact, the process of generating the 50 different initial solutions can be described as follows: 40 solutions will be generated by randomly selecting one from the set of non-dominated solutions, to which other security technologies are added until the construction-stopping criteria is satisfied. The 10 remaining solutions are generated randomly as described in Step2, in order to make a better diversity in the search space.

**5.2 Genetic algorithm:**

The genetic algorithm is a well-known meta-heuristic that was applied to a wide variety of single and multi-objective optimization problems. It is characterized of 2 main operators, the crossover and the mutation operators. The crossover operator is applied to generate children from a pair of parents selected from the current population. Each parent contributes by a portion of its genetic make-up to each child. And the mutation operator randomly changes a tiny amount of genetic information in each child.

The static phase of the proposed genetic algorithm can be described as follows: An initial population of 50 solutions is generated, similarly to the HS algorithm (refer to Step2). Then, with a probability of 90%, a two point crossover operator is applied to two randomly selected solutions from the population, to get two children. If the construction-stopping criterion is not satisfied for a child, the construction process will continue in the same way of the constructing the initial population process. Next, with a probability of 10%, the mutation operator is applied to each child. It consists on eliminating one of the security technology used by the solution and continue the construction process by the remaining security technologies. Finally, a child is added to the population if it is not dominated by any existing solution and if it is added, all the solutions dominated by it will be eliminated. This process stops if there is no improvement of the population for 50 successive iterations. The final population will constitute the set of non-dominated solutions.

The dynamic phase is identical to the one of the HS algorithm.

## 6. COMPUTATIONAL RESULTS

In this section, the performances of the two approaches are verified for different problem sizes. To do so, the qualities of non-dominated solutions generated by the two techniques for different instances are evaluated according to the $C$ metric (coverage of two sets) that can be defined as follows [9]: Let A, B be two non-dominated solutions sets. The measure C maps the ordered pair (A, B) into the range [0, 1]:

$$C(A, B) = \frac{|\{b \in B / \exists a \in A : a \text{ dominates } b\}|}{|B|} \quad (12)$$

This metric calculates, for a non-dominated solutions set B, the percentage of solutions that are dominated by at least a solution of the non-dominated solutions set $A$.

When testing the two approaches, it is supposed that there are, initially, 25 vulnerabilities and 40 security technologies, where each security technology addresses specific vulnerabilities (the $SV$ matrix). Then, in each time period new vulnerabilities and security technologies are added to the system and the $SV$ matrix is updated. It is supposed that a new vulnerability can be covered by an existing security technology or by a new one. In addition, a cost matrix is generated in such a way that more is the number of covered vulnerabilities by a security technology, higher is its relative cost. And finally, it is supposed that total cost allowed (*Cmax*) is 100.000 and the number of residual vulnerabilities (*Nvmin*) should not be more than 3.

The two algorithms are executed for 11 time periods and the results of the comparison are summarized in *Table1*, where the row *T* represents the time periods. The row *Size* is a pair (security, vulnerability) giving the number of the security technologies that may be used by the organization and the number of information system's vulnerabilities. The column *C(HS,GA)* presents the frequency by which the outcome of genetic algorithm is dominated by solutions generated by the HS algorithm. The column *C(GA,HS)* gives the frequency by which the outcome of the HS algorithm is dominated by solutions generate by the GA. And the row *Common Solutions* gives the number of similar solutions found by the two algorithms.

To detail the information given by table1 we take as example the time period 1. There are 35 vulnerabilities in the information system and the organization has to select its security plan among 55 security technologies. The results generated by the 2 algorithms indicate that 33% of the solutions generated by the GA are dominated by at least a solution generated by the HS algorithm. Whereas, there is no solution generated by the HS algorithm dominated by the non-dominated solutions of the GA. And there are 2 common solutions generated by the two algorithms.

The results presented in Table1 indicate that the HS algorithm generates in most times better results than the GA. In fact, among the 11 time periods, the C measure value was in the favor of the HS algorithm for 8 times against once for the GA and 2 equalities.

Table1: Comparison of the two approaches in term of C measure values

| Time period | Size | C(HS,GA) | C(GA,HS) | Common Solutions |
|---|---|---|---|---|
| 0 | (25, 40) | 25% | 0% | 3 |
| 1 | (35, 55) | 33% | 0% | 2 |
| 2 | (45, 70) | 75% | 25% | 0 |
| 3 | (55, 85) | 50% | 0% | 2 |
| 4 | (65, 100) | 0% | 0% | 4 |
| 5 | (75, 115) | 0% | 50% | 2 |
| 6 | (85, 130) | 50% | 0% | 2 |
| 7 | (95, 145) | 100% | 0% | 0 |
| 8 | (105, 160) | 75% | 25% | 0 |
| 9 | (115, 175) | 50% | 0% | 2 |
| 10 | (125, 190) | 50% | 50% | 0 |

## 7. CONCLUSION

In this paper, the problem of securing information systems was studied as bi-objective problem where we have to minimize the information system's vulnerabilities with a minimum cost. This problem was defined and formulated as a dynamic resource allocation decision model in order to protect, in real time, the organizations from the attacks frequently occurring. To solve this problem, a harmony search algorithm and a genetic algorithm were proposed. A comparison of the two approaches, according to the C measure was established. It indicates that the HS algorithm gives better results in most time periods of the optimization process.